# Linear and non-linear optical response of $Mg_xZn_{1-x}O$: A Density Functional study

G. Murtaza[1], Iftikhar Ahmad[1, *], B. Amin[1], A. Afaq[2], F. Ghafoor[3], A. Benamrani[4],

1. Department of Physics and Astronomy, Hazara University, Mansehra, Pakistan
2. Center for Solid State Physics, University of the Punjab, Lahore, Pakistan
3. Center for Quantum Physics, COMSATS Institute of Information Technology, Islamabad, Pakistan
4. Optoelectronics and Compounds Laboratory, Physics Department, Ferhat Abbes University, Serif 19000, Algeria

## ABSTRACT

Total and partial density of states, frequency dependent complex refractive index including extinction coefficient, optical conductivity and transmission of $Mg_xZn_{1-x}O$ ($0 \leq x \leq 1$) in rocksalt and wurtzite phases are calculated using full potential linearized augmented plane wave (FP-LAPW) method. The real part of refractive index decreases while the extinction coefficient, optical conductivity and transmission for rocksalt phase increases with the increase in Mg concentration. In wurtzite phase, ordinary and extraordinary indices decreases while extinction coefficient, optical conductivity and transmission increases in parallel as well as perpendicular to c-axis with the increase in the Mg concentration.

*     Corresponding author's email: ahma5532@gmail.com*

## 1. Introduction

ZnO is a key semiconductor compound with a wide bandgap, large binding energy and high chemical, mechanical and thermal stability [1]. It is intensively studied due to its numerous applications in sensors, transducers, catalysts, laser diodes, light emitting diodes, solar cells, heat mirrors, transparent electrodes, varistors, and surface acoustic wave devices [2-10]. While, MgO is another promising insulating metal oxide with various technological applications in different fields due to its very wide bandgap as well as thermal and chemical stability [11]. It is used as a substrate [12] and catalyst [13]. It is also used in superconductor products [14], thermal shock transducers, protecting layers for AC plasma display panels [15] and spintronics devices [16].

Mg doped ZnO is frequently used in optoelectronic and magneto-electronic devices [17-22]. $Mg_xZn_{1-x}O$ in wurtzite and cubic rocksalt form has been grown experimentally using different experimental techniques [23-31]. Its structural, electronic, optical and thermal properties have been theoretically investigated by different ab-initio methods [22, 34-40]. But so far, no theoretical work has been reported on the refractive index optical conductivity and transmission of $Mg_xZn_{1-x}O$ ($0 \leq x \leq 1$) alloys using density functional calculation.

In the present work we extend the theoretical work of Amrani et al. [33] and Schleife et al.[40] to the ground state optical response of $Mg_xZn_{1-x}O$ ($0 \leq x \leq 1$) in RS(B1) and WZ(B4) respectively, using Full- Potential Linearized Augmented Plane Wave Method with Perdew, Burke, Ernzerhof potential [41] in the generalized gradient approximation. In optical response we discussed in details the optical properties; refractive index including extinction coefficient, optical conductivity and transmission of the alloy.

## 2. Theory and calculations

Optical properties provide useful information about the internal structure of a material. Wide range of physical properties of a ternary compound makes it more suitable than a binary compound for many technological applications. Keeping this in mind we investigate the refractive index, optical conductivity and transmission of $Mg_xZn_{1-x}O$ ($0 \leq x \leq 1$). The refractive indices of ternary oxides are important optical design parameters, e.g. for distributed Bragg reflectors (DBRs) in vertical cavity surface emitting lasers [42].

The complex refractive index is given by:

$$\tilde{n} = n(\omega) + ik(\omega) = \sqrt{\varepsilon_1(\omega) + i\varepsilon_2(\omega)}$$

where $n(\omega)$ represents the real part of the refractive index and $k(\omega)$ is the extinction coefficient or attenuation index. $n(\omega)$ and $k(\omega)$ of a material can be evaluated by the real and imaginary parts of dielectric function, $\varepsilon(\omega)$, using the following relations:

$$n(\omega) = \frac{1}{\sqrt{2}}\left[\{\varepsilon_1(\omega)^2 + \varepsilon_2(\omega)^2\}^{1/2} + \varepsilon_1(\omega)\right]^{1/2} \quad (1)$$

$$k(\omega) = \frac{1}{\sqrt{2}}\left[\{\varepsilon_1(\omega)^2 + \varepsilon_2(\omega)^2\}^{1/2} - \varepsilon_1(\omega)\right]^{1/2} \quad (2)$$

Where the real and imaginary parts of the dielectric function are calculated using the following equations [43, 44]:

$$\varepsilon_2(\omega) = \frac{8}{2\pi\omega^2}\sum |P_{n\acute{n}}(k)|^2 \frac{dS_k}{\nabla\omega_{n\acute{n}}(k)} \quad (3)$$

$$\varepsilon_1(\acute{\omega}) = \frac{2}{\pi} P \int_0^\infty \frac{\acute{\omega}\, \varepsilon_{2(\acute{\omega})}}{\acute{\omega}^2 - \omega^2} \, d\acute{\omega} \tag{4}$$

where $P_{n\acute{n}}(k)$ is the dipole matrix element between initial and final states, $S_k$ is an energy surface with constant value, $\omega_{n\acute{n}}(k)$ is energy difference between two states and p denotes the principal part of the integral.

The frequency dependent optical conductivity is given by;

$$\sigma(\omega) = {2W_{ev} \hbar \omega}\Big/{\vec{E}_0} \tag{5}$$

where $W_{ev}$ is transition probability per unit time.

Transmission is calculated using the following relation [45]:

$$\alpha = 2.303 \times \ln(1/\text{T}) \tag{6}$$

DFT has become an unchallenging "standard model" in material science due to its strong predictive power and successful results with wide applicability. We used DFT with the generalized gradient approximation [41] in the wien2k package to solve Kohan Sham equations using full potential linear augmented plane wave (FPLAPW) technique [46]. For the crystal potential, muffin-tin model was assumed. The unit cell was divided into two parts, within and outside the muffintin. The electrons density within the muffin-tin is core and outside is valance. The core electrons were treated relativistically, whereas the valence electrons were treated non-relativistically. In both regions of the unit cell, different basis set were used to expand the wave function. A linear combination of radial solution of the Schrödinger equation times the spherical harmonics was used for the inside of the non-overlapping spheres of muffin-tin radius ($R_{MT}$), around each atom. While plane wave basis set was used in the interstitial region. $R_{MT}$ was chosen in such a way that there was no charge leakage from the core and total energy convergence was ensured. $R_{MT}$ values of *1.87, 2.02, 1.79* a.u. were used for Mg, Zn and O. Inside the atomic spheres maximum value of angular momentum $l_{max} = 10$, was taken for the wave function expansion. For wave function in the interstitial region the plane wave cut-off value of K$max$ = 7/ R$_{MT}$ for binaries MgO and ZnO while *10* for ternaries Mg$_{0.25}$Zn$_{0.75}$O, Mg$_{0.50}$Zn$_{0.50}$O, Mg$_{0.75}$Zn$_{0.25}$O was used. A mesh of *4000* k-points for the Mg$_x$Zn$_{1-x}$O ($0 \leq x \leq 1$) in RS (B1) and *550* for WZ (B4) were taken for the Brillouin zone integrations in the corresponding irreducible wedge to calculate optical properties.

## 3. Results and Discussions:

### 3.1. Density of states:

Total and partial density of states for Mg$_x$Zn$_{1-x}$O ($0 \leq x \leq 1$) in RS and WZ phases are calculated and presented in Fig. 1. In the figure the density of states (DOS) for WZ phase are plotted, while due to the similar trend it is not shown for RS. In pure ZnO, the upper part of the valence band (VB) is mainly composed of Zn-$3$d states while the lower part of this band is occupied by O-2p states along with a noticeable contribution from the hybridization of Zn-$3$d and O-$2$p states. The mixing of valence states above the Fermi level forms the conduction band. For all the ternary alloys the distribution of the states are almost the same as of pure ZnO except Mg-$3$p states, which also contribute to CB. The figure also reveals that for pure MgO the major contribution to the VB and lower part of CB comes from O-$2$p states, while upper part of CB is mainly composed of Mg-$3$p states. The calculated DOS for the both phases of ZnO and MgO are in agreement with Ref. [35]. The trend in the DOS in the RS and the WZ is same however there are some small differences i.e. DOS in of the WZ are higher than the RS.

### 3.2. Complex refractive index

For accurate modeling and designing of optical materials for devices, knowledge of the dispersion of the refractive indices of the medium is necessary. Rocksalt (RS) materials have cubic symmetry and therefore have isotropic properties, while wurtzite structures (WZ) lack cubic symmetry and therefore have anisotropic optical properties. The anisotropy results in uniaxial birefringence in wurtzite compounds. The one in the parallel direction to the c-axis is called ordinary ($n_o$) while the perpendicular one is called extraordinary ($n_e$) [5].

The calculated frequency dependent refractive index for $Mg_xZn_{1-x}O$, at *x = 0, 0.25, 0.50, 0.75,* and *1.0.* is shown in Fig. 3. The critical points (zero frequency limits) of refractive indices at *x = 0, 0.25, 0.50, 0.75* and *1.0* are plotted against Mg concentration in Fig. 2 for rock salt (RS) and wurtzite (WZ) structures. It is clear from the figure that the refractive index for each structure decreases with the increase in magnesium concentration in the ZnO crystal. It is also clear from the plot that the refractive index for rock salt is higher than that for the wurtzite.

In rock salt phase, the zero frequency limit of frequency dependent refractive index for pure ZnO is *1.95*. It increases from *1.95* with the increase in energy and reaches to its maximum value (peak value) of *2.37* at *3.58* eV. On further increase in the incident photon energy beyond the peak value the refractive index of pure ZnO dissipates with somewhat oscillations. It falls below *1* in the range *16.31-47.58* eV except for *29.73-31.06* eV. For the wurtzite phase of ZnO, the refractive index remains isotropic except in the range *10-20* eV, with the zero frequency limits *1.797* for parallel and *1.808* for perpendicular to c-axis. The calculated refractive indexes at *0.4* eV are *1.80* and *1.81* corresponding to parallel and perpendicular to c-axis respectively which are closer to the experimental results *1.93* and *1.96* respectively [47]. Our calculated results for both, parallel and perpendicular parts of the refractive index are also in excellent agreement with the results of Yoshikawa and Adachi [48]. The refractive index for pure wurtzite ZnO falls below *1* in the range *14.92-47.52* eV for both axis. It is well established fact that, the group velocity of a wave packet in a medium travels faster than the speed of light (c) for refractive index lesser than *1*. Hence the group velocity is faster than c for rock salt ZnO in the range *16.31-47.58* eV except *29.73-31.06* eV and for wurtzite ZnO in the range *14.92-47.52* eV. In other words the group velocity shifts to negative domain and the material does not stay linear but becomes nonlinear for these energy ranges and hence becomes super luminous.

By increasing the Mg concentration in ZnO for $Mg_xZn_{1-x}O$, the peak value of the refractive index shifts towards higher energies for both the phases which is clear from Fig.2. When all the Zn atoms are replaced by Mg in the crystal, the peak value of the refractive index of the pure RS-MgO (*1.89*) occurs at *10.2* eV and WZ-MgO (*1.67*) at *6.1* eV. Our calculated ground state refractive index of RS-MgO, *1.55*, is closer to the room temperature experimental refractive index, *1.72* [49]. Overall, by the change in the Mg concentration in the ZnO crystal, we can obtain different refractive indexes. For zero frequency limits they range from *1.95* to *1.55* in RS and, and for peak values they lie between *2.37* and *1.89*, with the shift of energy from lower towards higher. Similarly, in WZ phase it varies from *1.797* to *1.428* parallel and *1.808* to *1.438* perpendicular to the c-axis.

Extinction coefficient against energy is also shown in Fig. 2 for both RS and WZ phases. Critical values for the extinction coefficient of $Mg_xZn_{1-x}O$, at *x = 0, 0.25, 0.50, 0.75, 1.0* are; *1.6* eV, *2.05* eV, *2.55* eV, *3.1* eV, *5.1* eV for RS phase while *1.2* eV, *1.7* eV, *2.06* eV, *2.8* eV, *5.8* eV for parallel as well as perpendicular to c-axis of the WZ phase. In pure RS-ZnO, the prominent peaks are at *12.4* eV and *16* eV. When we replace Zn by Mg in the RS-ZnO crystal, the positions of the peaks shifts towards higher energy values until all the Zn are replaced by Mg and it is noted that (for RS-MgO) the prominent peaks are at *14.2* eV, *16.4* eV and *19.9* eV. It is further noted that the maximum peak in RS-ZnO is at *16.06* eV, while in RS-MgO it occurs at *19.87* eV.

So by increasing Mg concentration in the RS-ZnO crystal, the overall response of the material to the applied energy field shifts towards higher energy in the ultraviolet region of the spectrum.

In WZ phase, the same trend is observed for the peak values of the extinction coefficient parallel to the c-axis from *14.84* eV to *17.62* eV while in the perpendicular part it decreases with the increase in the Mg concentration, from *13.67* eV to *12.20* eV. The anisotropy found in the parallel and perpendicular parts of the extinction coefficient for WZ-ZnO is in the range *6.8* eV to *17.5* eV and in WZ-MgO *6.89* eV to *29* eV, while the ternary alloys are isotropic for most of the energies. The calculated refractive index and extinction coefficient shows that $Mg_xZn_{1-x}O$ in both phases is suitable for optoelectronic applications in the ultraviolet region of the spectrum.

### 3.3. Optical conductivity

Optical conductivity, $\sigma(\omega)$, of $Mg_xZn_{1-x}O$ ($0 \leq x \leq 1$) is shown in Fig. 4 for both phases. The critical points for $Mg_xZn_{1-x}O$ at $x = 0, 0.25, 0.50, 0.75, 1.0$ are *2.6* eV, *3.05* eV, *3.5* eV, *4.12* eV, *5.5* eV for RS phase and 1.3 eV, 2.06 eV, 2.65 eV, 3.68 eV, 5.80 eV for WZ phase. In pure RS-ZnO, there are three significant peaks at *12.6* eV, *15* eV and *31* eV with a maximum value of 6882.79 $(\Omega.\text{cm})^{-1}$ at 14.84 eV. The most interesting peak is the middle one, around 15 eV, which increases with the increase in the Mg concentration till 25 % and on further increase it starts decreasing. The peak around 20 eV is also appealing, as it increases with the increase in the Mg concentration and becomes prominent for pure RS-MgO. It is further noted that the maximum peak value of conductivity shifts from *14.84* eV to *19.6* eV.

Similar trend are also observed for the optical conductivity parallel as well as perpendicular to c-axis. The maximum conductivity for WZ-ZnO is *6201.52* $(\Omega.\text{cm})^{-1}$ at *13.18* eV for parallel to c-axis and *6763* $(\Omega.\text{cm})^{-1}$ at *13.59* eV for perpendicular to c-axis. While in the

WZ-MgO, maximum conductivity parallel to c-axis is *4312.83* ($\Omega \cdot cm$)$^{-1}$ at *16.37* eV and perpendicular to c-axis is *5188.04* ($\Omega \cdot cm$)$^{-1}$ at *16.34* eV. Hence with the increase of Mg concentration, the peak value of the optical conductivity decreases and shifts towards higher energies for both parallel and perpendicular components to c-axis. The anisotropy observed in optical conductivity of WZ-ZnO is in the range *11.4* eV to *30* eV and WZ-MgO is *6.89* eV to *29* eV, while the ternary compound $Mg_xZn_{1-x}O$ is almost isotropic.

### 3.4. Transmission

Transmission spectra of $Mg_xZn_{1-x}O$, for both phases, are calculated for the incident photon energies up to *7.0* eV using Eq. 6 and is plotted in Fig. 5. For RS-ZnO the transmission is almost 100% in the lower energies below 1 eV and then it starts decreasing exponentially and dies to zero near the absorption edge of the material. Similar trends are also observed for RS-MgO as well as their ternary RS-$Mg_{0.50}Zn_{0.50}O$ compounds. For uniaxial WZ-$Mg_xZn_{1-x}O$, the material remains isotropic in the range *0.0* eV – *7.0* eV for both transmissions parallel as well as perpendicular to c-axis. Transmission parallel to c-axis is plotted in Fig. 5. From plot it is clear that the exponential decrease in WZ-ZnO is faster than RS-ZnO but the transmission edge comes beyond RS-ZnO. Similar trend is also observed for WZ-$Mg_{0.50}Zn_{0.50}O$. It is further noted from the figure that the transmission for both phases of MgO remains similar for most energies, while the transmission in WZ-MgO decays earlier than RS-MgO. For both phases the transmission increases with the Mg concentration in the ZnO crystal and the transmission edge shifts towards higher energies in the ultraviolet region of the spectrum. The results obtained from the present density functional calculations are in excellent agreement with the experimentally measured transmission of WZ-ZnO [31] and WZ-$Mg_xZn_{1-x}O$ [30] and are also in agreement with the experimental results of $Mg_xZn_{1-x}O$ in rocksalt and wurtzite [31].

## 4. Conclusion

Density of states, refraction, optical conductivity and transmission of $Mg_xZn_{1-x}O$ ($0 \leq x \leq 1$) in rocksalt and wurtzite structures are theoretically investigated by FP-LAPW method. The calculated refractive index show that the material stays linear for energies lesser than 17eV in RS and 15 eV in WZ and becomes nonlinear beyond this. The increase of $x$ in $Mg_xZn_{1-x}O$ crystal, suppresses the overall refraction and conductivity except for the sharp peak value around *19.90* eV in RS and *16.3* eV in WZ, which enhances. Transmission of the alloy in both phases increases with the increase in Mg concentration. The high response of the compound in the energy range 10 eV to 20 eV confirms the importance of the material in the UV optoelectronic devices.

**Acknowledgements**

Prof. Dr. Keith Prisbrey, MSE, University of Idaho and Prof. Dr. Nazma Ikram, Ex. Director, Center for Solid State Physics, Punjab University are highly acknowledged for their valuable suggestions.

**Figure Caption**

**Fig. 1:** Total and partial density of states for Mg$_x$Zn$_{1-x}$O in wurtzite phase.

**Fig. 2:** Real and imaginary parts, $n(\omega)$ and $k(\omega)$, of frequency dependent refractive index of Mg$_x$Zn$_{1-x}$O ($0 \leq x \leq 1$)

**Fig. 3:** Static refractive index (zero frequency limits) of Mg$_x$Zn$_{1-x}$O ($0 \leq x \leq 1$)

**Fig. 4:** Frequency dependent optical conductivity $\sigma(\omega)$ of Mg$_x$Zn$_{1-x}$O ($0 \leq x \leq 1$)

**Fig. 5:** Transmission spectra of Mg$_x$Zn$_{1-x}$O ($0 \leq x \leq 1$)

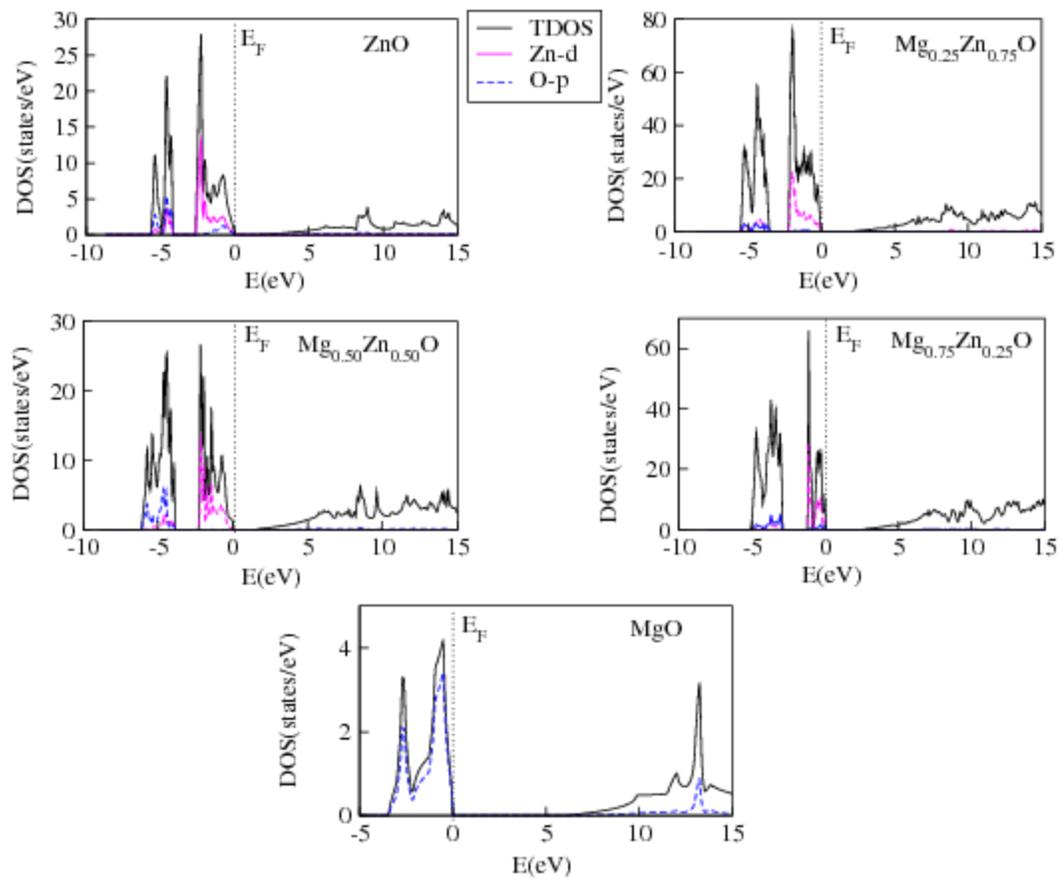

**Figure 1**

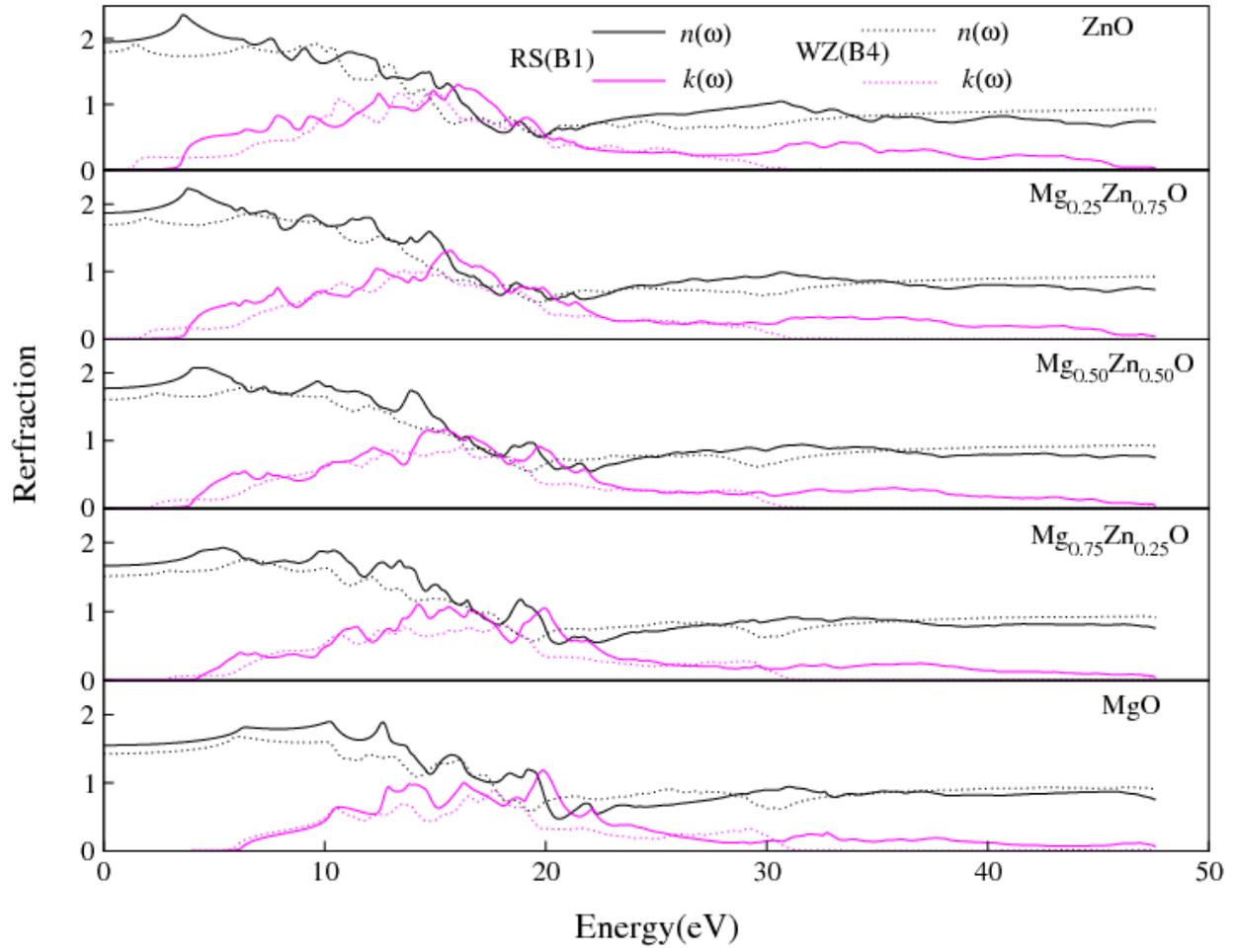

**Figure 2**

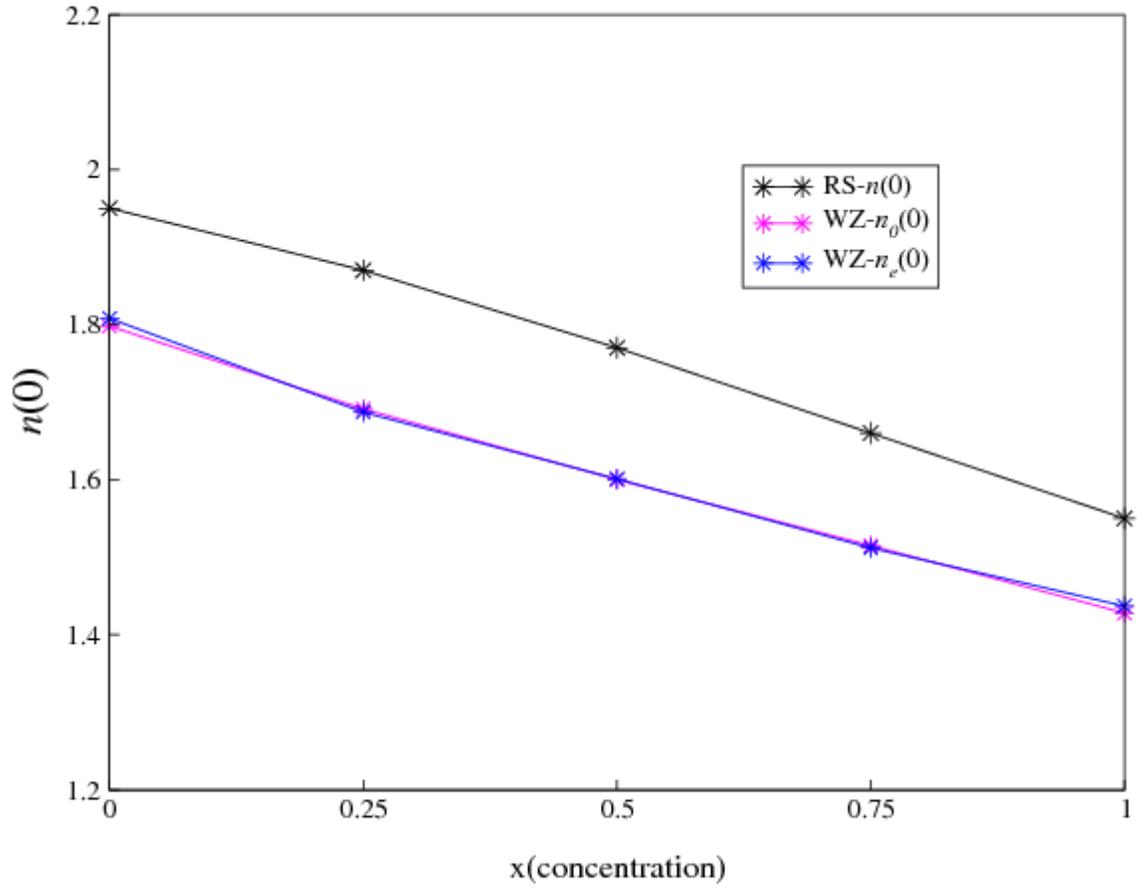

**Figure 3**

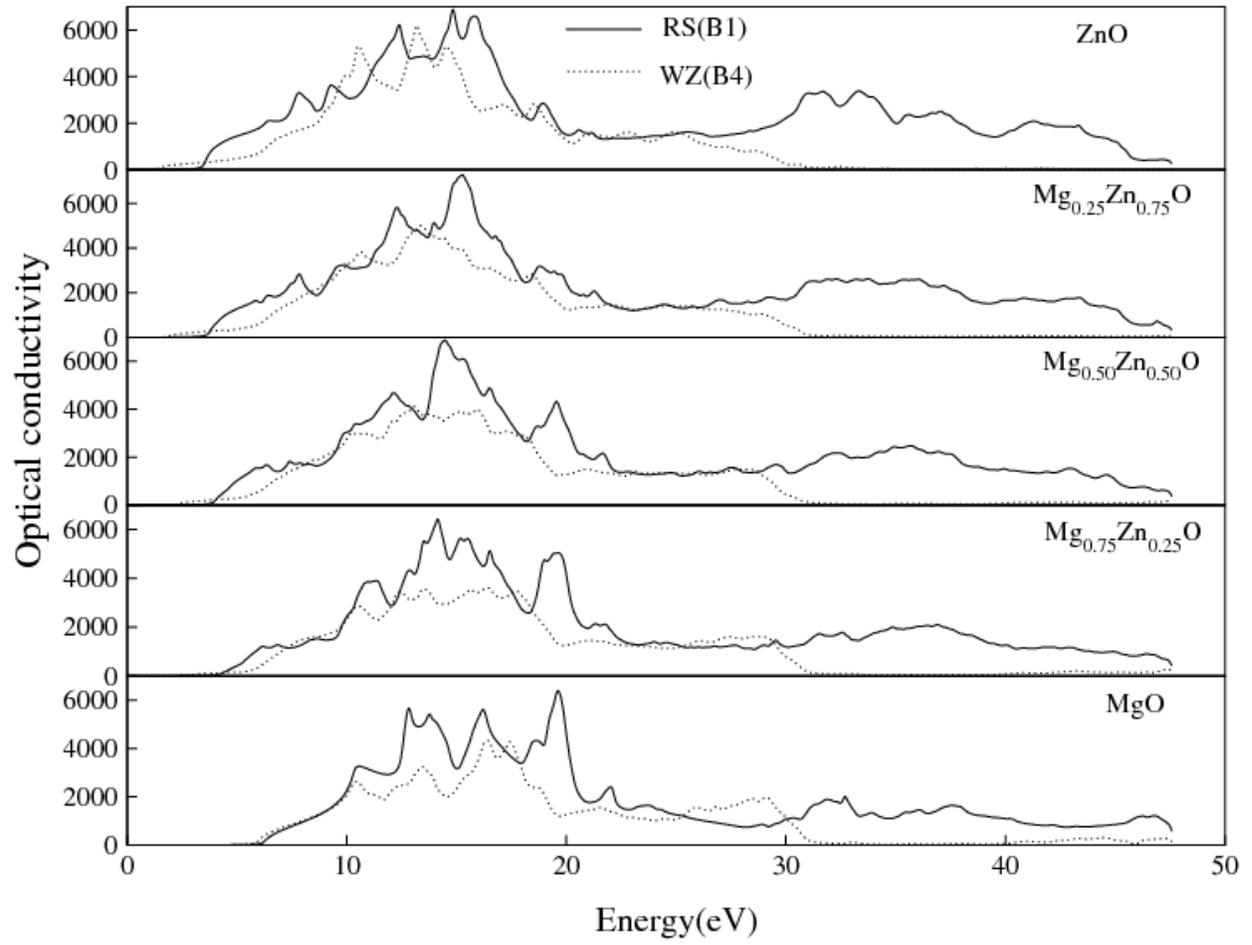

**Figure 4**

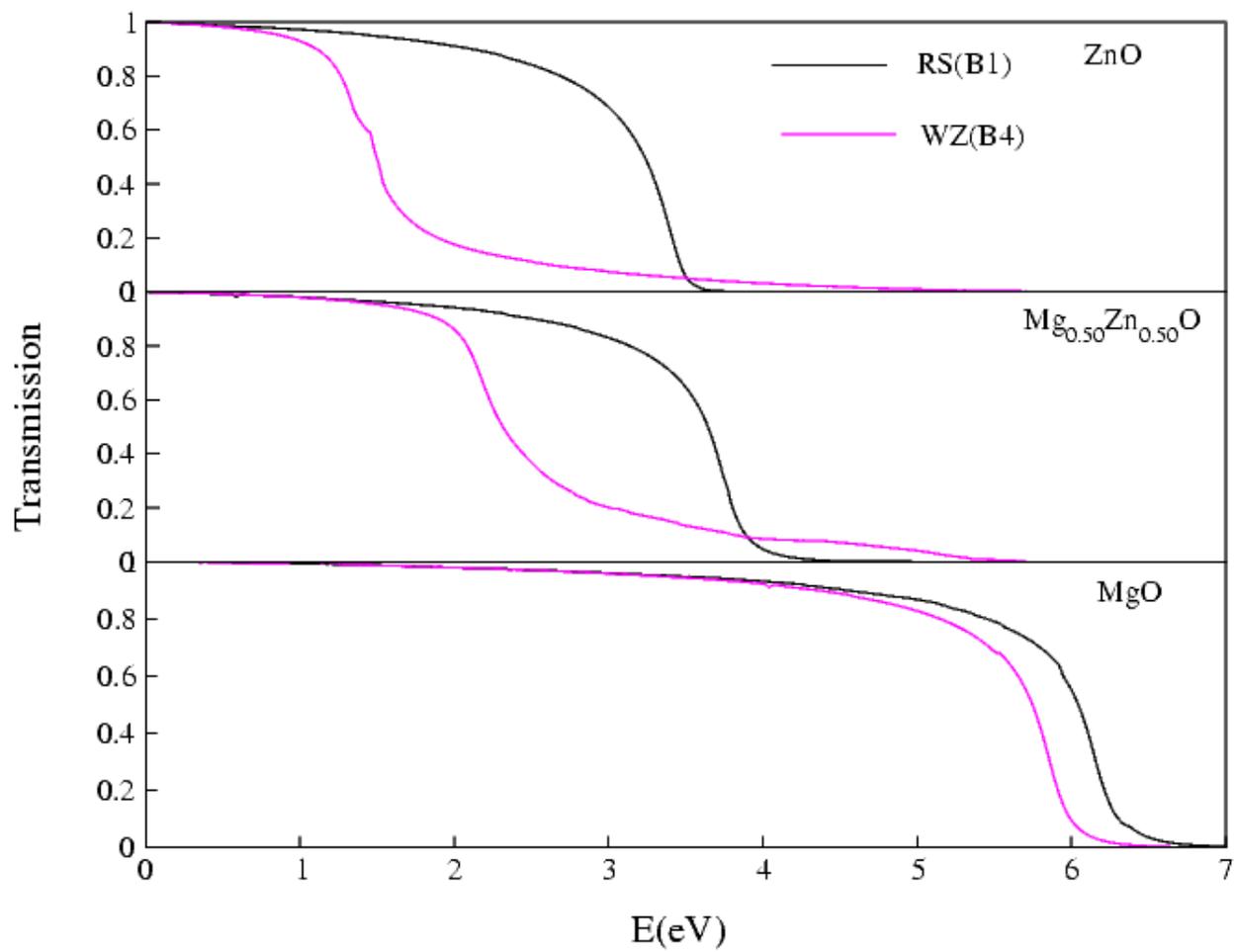

Figure 5